\documentclass[11pt]{article}
 \usepackage{mathtools}
 
 \usepackage[iso-8859-7]{inputenc}
 \usepackage{epsfig}
 \usepackage{graphicx}
 \usepackage{epstopdf}
\usepackage{amsfonts}
 \usepackage{amssymb}
 \usepackage{amsthm}
 \usepackage{amsmath}
 \usepackage{multirow}
 \usepackage{color}

  \newcommand{\eps}{{\epsilon}}
 \newcommand{\beq}{\begin{equation}}
 \newcommand{\eeq}{\end{equation}}
 \newcommand{\bea}{\begin{eqnarray}}
 \newcommand{\eea}{\end{eqnarray}}
 
 \newcommand{\gsim}{\lower.7ex\hbox{$\;\stackrel{\textstyle>}{\sim}\;$}}
 \newcommand{\lsim}{\lower.7ex\hbox{$\;\stackrel{\textstyle<}{\sim}\;$}}
 \newcommand{\be}{\begin{equation}}
 \newcommand{\ee}{\end{equation}}
 \newcommand{\ba}{\begin{eqnarray}}
 \newcommand{\ea}{\end{eqnarray}}

 \usepackage{cite}
 \usepackage{float}
 \usepackage{amsmath} 
 \usepackage{amssymb}   
 \usepackage{amsthm} 
 \usepackage{geometry}
 \usepackage{graphicx} 
 \usepackage{multirow} 
\begin{document} 
 	\thispagestyle{empty}
 	\begin{titlepage}
 		\vspace*{0.7cm}
 		\begin{center}
 			{\Large {\bf Note on de Sitter vacua from perturbative and non-perturbative dynamics	in type IIB/F-theory compactifications
			}}
 			\\[12mm]
 			Vasileios Basiouris~\footnote{E-mail: \texttt{v.basiouris@uoi.gr}},
 			George K. Leontaris~\footnote{E-mail: \texttt{leonta@uoi.gr}}
 		\end{center}
 		\vspace*{0.50cm}
 		\centerline{\it
 			Physics Department, University of Ioannina}
 		\centerline{\it 45110, Ioannina, 	Greece}
 		\vspace*{1.20cm}
 \begin{abstract}

The properties of the effective scalar potential are studied  in the framework of type IIB string theory, taking into account  
perturbative and non-perturbative corrections. The former  modify the K\"ahler potential and include $\alpha'$ and logarithmic 
corrections generated when intersecting $D7$ branes are part of the internal geometric configuration. The latter add exponentially 
suppressed K\"ahler moduli dependent terms to the fluxed  superpotential.  The possibility of partial elimination of such terms 
which may happen  for particular choices of world volume fluxes is also taken into account. That being the case, a simple set up of three 
K\"ahler moduli is  considered  in the large volume regime, where only one of  them  is assumed  to induce  non-perturbative corrections.  
It is found  that the shape of the F-term potential crucially depends on  the parametric space associated with the perturbative sector 
and the volume modulus. De Sitter vacua can be obtained  by implementing  one of the standard mechanisms, i.e., either relying on D-terms  
related to $U(1)$ symmetries associated with the  $D7$ branes, or introducing $\overline{D3}$ branes. In general it is observed that the 
 combined effects of  non-perturbative dynamics and the recently introduced logarithmic corrections lead to an effective scalar potential 
 displaying interesting cosmological and phenomenological properties.

\end{abstract}

\end{titlepage}

\section{Introduction}

Recent swampland conjectures~\cite{Vafa:2005ui,Obied:2018sgi,Agrawal:2018own} have sparked an interesting discourse regarding the nature of 
string theory vacua~\footnote{For related reviews and further references see~\cite{Palti:2019pca} and~\cite{Danielsson:2018ztv}.}. 
The ``underlying essence'' of these 
 hypotheses  is that  the string landscape does not contain stable de Sitter (dS) vacua. However, 
it was asserted by several authors, that dS minima are in principle possible in string theory, when the physics implications of 
  perturbative and non-perturbative   dynamics are taken into account.  Indeed, quantum corrections in  string theory  play a key 
  r\^ole in shaping the final  form of the effective theory.   In many cases, however,
 a straightforward approach to incorporate these effects  ends up to an  effective theory with anti de-Sitter (AdS)  vacuum  and thereby 
 an appropriate mechanism is necessary  to uplift it to a dS vacuum.   Generally, the last two decades or so, in order 
 to  tackle the  problems  of moduli stabilisation and generate dS vacua, various sources of corrections were considered including 
 non-perturbative  objects such as D-branes,  string loop effects etc.  A great deal of endeavours  to  stabilise the K\"ahler moduli and 
 construct  stringy dS  vacua  has been focused  on non-perturbative corrections~\cite{Kachru:2003aw} and large   volume  
 compactifications~\cite{Conlon:2005ki} generated by Euclidean D3 branes~\footnote{For recent work on KKLT see~\cite{Polchinski:2015bea}-\cite{AbdusSalam:2020ywo} and for earlier contributions see~\cite{Abe:2005rx,Parameswaran:2006jh}.
 	For cases suggesting small ${\cal W}_0$ values see~\cite{Demirtas:2019sip,Carta:2019rhx}.}.  
 The former use  an uplift mechanism based on  anti-D3 branes ($\overline{D3}$ for short) to generate a dS vacuum.    
In the large volume  compactification scenaria  (LVS), (see \cite{Burgess:2020qsc}  for a recent perspective) leading order $\alpha'$
 perturbative corrections~\cite{Becker:2002nn}  were also  included in the K\"ahler potential and it has been argued that they are dominant
  at  large as  well as  small volumes. The present work will be in the 
  framework of type IIB string and F-theory compactifications  with D-branes and 
  fluxes~\footnote{For a general review regarding four-dimensional  compactifications with D-branes and fluxes see~\cite{Blumenhagen:2006ci}.}
 where such mechanisms are 
available.

 Recently, it was shown that the issues of moduli stabilisation  and  de Sitter vacua can be successfully resolved
	 with 	contributions arising only from
	perturbative string-loop corrections~\cite{Antoniadis:2019rkh}. 
	These are related to higher derivative terms in the effective string action which  generate  
	localised  Einstein-Hilbert terms
	and, in geometric configurations involving $D7$-branes, induce logarithmic corrections via string loop effects. 
It has also been shown that cosmological inflation is successfully implemented~\cite{Antoniadis:2020stf}.

	Quantum  corrections of this type are standard in the presence of D-branes and  
	were also studied in the past~\cite{Antoniadis:1998ax,Goldberger:2001tn}
	although in different contexts.
	Also, in~\cite{Ibanez:1999pw} it was shown that invariance of the effective classical action under
	$SL(2,R)$ transformations  implies logarithmic corrections to the K\"ahler potential which depend on 
	the untwisted K\"ahler moduli.
	Such contributions,  break the no-scale structure of  the K\"ahler potential and lead to an effective theory with 
	all K\"ahler moduli stabilised. Furthermore, D-term contributions related to the abelian symmetries 
	of the intersecting $D7$-branes, work as an uplift mechanism and
	ensure the existence of de Sitter vacua. 
	
	In the present work both perturbative logarithmic corrections in the K\"ahler potential
	as well as non-perturbative contributions to 
	 the superpotential  are combined,  and  stabilisation 
	 of the moduli fields and the  properties of the effective  potential are investigated. 
	With regard to the non-perturbative sector,  
	 an interesting and non-trivial  case will be  considered where only a subset of the K\"ahler moduli are 
	 	involved  in the superpotential  leaving the remaining degrees of freedom to be stabilised 
	 	by the quantum corrections breaking the no-scale structure of the K\"ahler potential. 
	 	For example, in the case of Euclidean instanton contributions, as 
	 described in~\cite{Bianchi:2011qh} (see also\cite{Duff:1996rs}), such a situation can occur if 
	the chosen world-volume fluxes lift certain fermionic zero-modes,
	 	whose presence would prevent the generation of non-perturbative superpotential terms.
As a working example in the present analysis, a composition of three K\"ahler 
moduli is considered with
only one of them  involved  in the non-perturbative part 
of the superpotential. Extensions to more than three moduli are straightforward, although the analysis becomes more involved
and will not be considered in this letter.   

 Section 2 begins with a short description of quantum corrections 
 where  the main focus is  on the logarithmic ones which prevail in a set up involving $D7$ branes.  The analysis of the minimisation conditions on the F-term potential  is presented in section 3 while in section 4   D-term
contributions from abelian factors related to the $D7$ brane configuration
are discussed. Then, the implementation of the uplift mechanism  through D-terms after some of the K\"ahler moduli obtain their critical values is explained. A summary of the work and the main  conclusions can be found 
in  section 5.

\section{Quantum corrections} 	

The aim of this work is to propose a solution to the moduli stabilisation problem in IIB string theory, combining both
 perturbative and non-perturbative corrections. Furthermore, the necessary conditions for the existence of  de Sitter minima
  will be investigated within the same string theory framework. 
The notation for  various  fields used in the subsequent analysis  is as follows: The dilaton and Kalb-Ramond fields are denoted 
with $\phi$ and $B_{2}$ respectively while  the various $p$-form potentials  with $ C_p,\, p=0,2,4$. 
The $C_0$ potential and the dilaton field $\phi$  are combined in the usual axion-dilaton combination:
\[S = C_0+{ i\,e}^{-\phi} \equiv C_0+\frac{i}{ g_s}~\cdot \]
Finally,  $z_a,\; a=1,2,3,...$ stand for the complex structure (CS) moduli  and  $T_i,\,i=1,2,3,\dots$ for the K\"ahler fields.
The fluxed induced  superpotential, $ {{\cal W}_0}$, at the classical level is~\cite{Gukov:1999ya}
\begin{eqnarray} 
{\cal W}_0=  \int\, G_3\wedge { \Omega}( z_a)~,\label{SupW}
\end{eqnarray} 
with $\Omega( z_a)$  being the holomorphic (3,0)-form and  $ G_3:=  { F_3}-{ S}\, { H_3}$, where the field
strengths are $ F_p:= d\,{ C_{p-1}}, H_3:=d B_{2}$.
\noindent 
The perturbative superpotential $ {{\cal W}_0}$ is a holomorphic function and depends on the 
axion-dilaton modulus $S$, and the CS moduli $z_a$. Thus, at the classical level, the supersymmetric 
conditions, ${\cal D}_{ z_a}{\cal W}_0=0$ and ${\cal D}_{ S}{\cal W}_0=0$ 
fix  the moduli $z_a, S$, however, 
the K\"ahler moduli remain completely 
undetermined. At the same order, 
the   K\"ahler potential   depends 
logarithmically on the various  fields, including the K\"ahler moduli 
\begin{eqnarray} 
{\cal K}_0=-\sum_{i = 1}^3 \ln(-i({ T_i-\bar{T_i}}))
- \ln(-i({ S-\bar{S}}))-\ln(-i\int{\Omega}\wedge {\bar{\Omega}})~\cdot\label{KahlerP}
\end{eqnarray}
Then, the effective potential is computed  using the standard formula
\begin{eqnarray} 
{ V}_{\rm eff }&=&e^{\cal K}\left(\sum_{I,J} {\cal D}_I{\cal W}_0{\cal K}^{I\bar J}{\cal D}_{\bar J}\overline{{\cal W}}_0-3 |{\cal W}_0|^2\right)
~\cdot\label{KahlerV}
\end{eqnarray}
In the absence of any radiative corrections, 
the latter vanishes identically  due to supersymmetric conditions and the no scale structure of the K\"ahler potential.
Hence, it is readily inferred  that in order to stabilise  the K\"ahler moduli it is necessary   to go beyond the classical level. In fact, when 
quantum corrections are included they break the no-scale structure of
the K\"ahler potential and presumably a non-vanishing contribution in the scalar potential, i.e. ${ V}_{\rm eff }\ne 0$, is feasible.

As already stated, in the quest for a stable dS minimum in effective string theories, the r\^ole of perturbative as well
 as non-perturbative corrections will be analysed. Furthermore it should be  mentioned that this work takes place in the framework of
 type IIB string theory compactified on a 6-d Calabi-Yau (CY) manifold $ {\cal X}_6$, and the 10-d space is ${\cal M}_4\times {\cal X}_6$.
 The subsequent computations are assumed in the   context of type IIB string theory compactified on the $T^6/Z_N$ orbifold limit of the CY space.
Furthermore,  a geometric configuration consisting  of three intersecting $D7$ branes is considered,
while the internal  volume ${\cal V}$  is expressed in terms of the imaginary parts  $v^i$ (the two-cycle volumes)
of the K\"ahler moduli 
\be 
{\cal V}= \frac 16 k_{ijk} v^iv^jv^k,\;  v^i= -{\rm Im}(T^i)~,
\ee 
where $ k_{ijk}$ are intersection numbers. The $v^i$  are related to 4-cycle volumes $\tau_i$ as follows:
\be 
\tau_i =\frac 12 k_{ijk}v^jv^k~.\label{4to2mod}
\ee 
In the present case it is simply assumed that ${\cal V}=v^1v^2v^3$ or,
in terms of the 4-cycle volumes $\tau_i$'s:
\be 
{\cal V}=\sqrt{\tau_1\tau_2\tau_3}~.\label{Vtau123}
\ee 
After these preliminaries, in the remaining of this section the various types of corrections
will be presented. 

\noindent 
 Starting with  non-perturbative corrections of the superpotential, 
in principle,  all three K\"ahler moduli considered in this model may contribute. 
In this case the superpotential takes the form 
\be  
{\cal W}= {\cal W}_0+ \sum_{k=1}^3 A_k e^{i a_k\rho_k}~,\label{Wnp3}
\ee 
In the above formula,  $\rho_k= b_k+i \tau_k$ where $b_k$ is associated 
with the RR $C_4$ form, $\tau_k$ is given by~(\ref{4to2mod}) and 
 $ {\cal W}_0=\int G_3\wedge \Omega$ is the tree-level superpotential in~(\ref{SupW}). The second term in the right-hand side of (\ref{Wnp3}) is the non-perturbative part~\cite{Witten:1996bn}.  The  constants $A_i$ in general depend on the complex structure moduli and the $a_i$ parameters are assumed
to be small (for example in the case of  gaugino condensation in an $SU(N)$, they are of the form $\frac{2\pi}{N}$). However, it maybe possible that  the choice of world-volume fluxes~\cite{Bianchi:2011qh}
allow  only some of the  K\"ahler  moduli fields to have non-vanishing non-perturbative (NP) contributions. In what follows,  it is supposed that there are no NP terms associated with $\tau_{2,3}$, and
   only the $\tau_1$ modulus induces a non-vanishing contribution in the NP part of the superpotential, thus
\be  
{\cal W}= {\cal W}_0+  A e^{i a\,\rho_1}~\cdot \label{Wnp1}
\ee 

\noindent 
Before proceeding to the next step, some comments are due with respect to (w.r.t.) the reliability of the instanton correction  and the specific choices in the subsequent analysis. This type of corrections originates from the presence of Euclidean D3-branes wrapping four-cycles in the base  of the compactification \cite{Witten:1996bn}. First of all, in order the supergravity approximation to be valid, the condition $\tau_1 \geq 1$ should be fulfilled. Two main reasons are in favor of this argument. First,  shrinking  one direction to small volume leads to highly curved K\"ahler cones or orbilfolds where the effective approximation is at stake. Second, the logarithmic correction~\cite{Antoniadis:1998ax} 
that has been  added in the K\"ahler potential requires large transverse directions $\tau_i$. We come back to this issue in section 3.2.

\noindent 
Next,  quantum corrections to the K\"ahler potential will be discussed,
starting with the ${\alpha}^{'3} $ contributions, which,  in the large volume 
limit  imply a redefinition of the dilaton field~\cite{Becker:2002nn}
\begin{eqnarray}
e^{-2\phi_4}&=&  e^{-2\phi_{10}}({\cal V} +\xi) 
= e^{-\frac 12\phi_{10}}\,(\hat{\cal V} +\hat{\xi})~\cdot \label{dilvol}
\end{eqnarray}
The last expression on the right-hand side of (\ref{dilvol}) holds in the  Einstein frame and the volume is written 
 in terms  of the imaginary parts  of the K\"ahler deformations $T^k$ as follows
 \begin{equation} {\cal V}=\frac{1}{3!}\kappa_{ijk}v^iv^jv^k,\;   v^k= -{\rm Im}({ T^k})=
\hat{ v}^k\,e^{\frac 12\phi_{10}}~\cdot
\end{equation}
The  modifications  in the K\"ahler potential 
correspond to a shift of the volume by a constant $\xi$  which is
determined in terms 
of the Euler characteristic
${\xi}=-\frac{\zeta(3)}{4(2\pi)^3}{\chi}$.

The origin of the second type of corrections comes from higher derivative terms 
which give rise to multigraviton scattering in string theory.    In type IIB theories, the leading terms appearing 
 in the 10-dimensional effective action 
 are proportional to $R^4$, where $R$ is the Riemann curvature. In theories with ${\cal N}=1$ sypersymmetry in 10 dimensions,
 the leading corrections already appear at order $R^2$. Here, the terms of interest to us are the $R^4$ couplings, which, 
after compactification to four dimensions ${\cal X}_{10}\to{\cal M}_4\times {\cal X}_6$, they induce a new Einstein-Hilbert (EH) term, 
 multiplied by the Euler characteristic of the manifold.
The  one-loop amplitude of the on-shell scattering involving four gravitons has been worked out 
in~\cite{Kiritsis:1997em,Green:1997di,Russo:1997mk,Antoniadis:1997eg, Antoniadis:2002tr,Antoniadis:2003sw,Berg:2005ja,Haack:2018ufg} 
where it has been shown that the ten-dimensional action reduces to
\begin{eqnarray} 
{\cal S }_{\rm grav}&=& \frac{1}{(2\pi)^7 \alpha'^4} \int\limits_{M_{4} \times {{\cal X}_6}} e^{-2\phi} {\cal R}_{(10)} - \frac{\chi}{(2\pi)^4 \alpha'} \int\limits_{M_{4}} \left(-2\zeta(3) e^{-2\phi}  \pm 4\zeta(2) \right) { R}_{(4)}\,,  
\label{IIBAction} 
\end{eqnarray} 
where $ { R}_{(4)}$ denotes the `reduced' Riemann tensor  in four dimensions, the $\pm$ signs refer to the type IIA/B theory respectively,
 and the Euler characteristic is defined as
\be 	
\chi= 	\frac{3}{4\pi^3}\int\limits_{{\cal X}_6} R\wedge R\wedge R~\cdot 
\label{Euchi}
\ee
From (\ref{IIBAction}), it is observed that  a localised EH  term~\footnote{The computations have been performed in the orbifold limit~\cite{Antoniadis:2019rkh}  
	and localisation occurs at the orbifold fixed points $p_i$. 
	These	points  correspond to the singularities where the Euler number is non-vanishing and  in general  
	$\chi= \sum_i\chi_{p_i}$. In this sense, the existence of the term ${\cal R}_{(4)}$ is associated with 
	these points, hence the term ``localised gravity''.} is generated with a coefficient proportional to   $\chi$ defined in~(\ref{Euchi}). Consequently it is inferred 	that this term is possible only in four dimensions.  In the geometry of the bulk space, the $R_{(4)}$ EH terms
 of (\ref{IIBAction}) correspond to vertices at points where  $\chi\ne 0$, and as such, they emit gravitons and Kaluza-Klein (KK) excitations in the six-dimensional space. 
Furthermore, in the presence of $D7$ branes which are an essential ingredient of the internal space configurations
in type IIB and F-theory, new types of quantum contributions emerge. It is  found thereby that  the exchange 
of closed string modes between the EH-vertices and $D7$ branes and $O7$-planes
give rise to logarithmic corrections. These take  the form~\cite{Antoniadis:2019rkh}
\begin{eqnarray}
\frac{4\zeta(2)}{(2\pi)^3}\chi \int_{M_4} \left(1-\sum_k e^{2\phi}T_k \ln(R^k_{\bot}/\mathtt{w})\right)\,R_{(4)}~.\label{allcor}
\end{eqnarray}
In the above, $T_k$ is the  tension of the $k^{th}$ 7-brane, $R_{\bot}$ stands for the size of 
the two-dimensional space transverse to the brane, and $\mathtt{w}$ is a `width' related to  an effective ultraviolet
cutoff  for the graviton KK modes propagating in the bulk~\cite{Antoniadis:2002tr}.

\section{The effective potential}

Assembling  the above ingredients  the effective potential is readily constructed. To facilitate the presentation, firstly, the F-term part 
will be discussed and
afterwards the contributions from D-terms
will be included.

\subsection{The K\"ahler potential and the superpotential}

Considering  a geometric configuration of three intersecting  $D7$ branes,  the  corrections (\ref{allcor}) imply that the	contributions involving the  K\"ahler  moduli  are of the form~\cite{Antoniadis:2019rkh} 
${ \delta } =\xi+ \sum_{k=1}^3{ \eta_k} \ln ({ \tau_k})$.  For simplicity 
it is assumed that all $D7$ branes have the same tension $T=e^{-\phi}T_0$, 
so the order one coefficients   $\eta_k$ and $\xi$ are given by
\be 
\eta_k\equiv \eta =-\frac{1}{2} g_sT_0\;\;;\;\; \xi=-\frac{\chi}{4} \times 
\begin{cases}
\frac{\pi^2}{3}g_s^2\quad {\rm for\ orbifolds}
\\[3pt]
\zeta(3)\;\quad {\rm for\ smooth\ CY}
\end{cases}\label{etaxi}
\ee 
and  $\tau_k$ are defined in (\ref{4to2mod}).
As can be observed in~(\ref{etaxi}), tree-level contributions for the orbifold have  not  been included, since,
 according to the  arguments presented by the authors of~\cite{Antoniadis:2002tr},  
in this case the $\zeta(3)\chi$ correction to EH term vanishes~\footnote{Of course, one may not adopt the reasoning of the authors 
	of reference~\cite{Antoniadis:2002tr} and  argue that
since the  orbifold	does have a non-vanishing Euler number, there is also a non-zero tree-level contribution in this case too. Nevertheless, this contribution  does not modify  the present analysis.}. It should also be pointed out that
one could consider subleading 1-loop corrections to $\xi$ in the smooth CY case.

\begin{figure}[H]
	\centering
	\includegraphics[scale=.8]{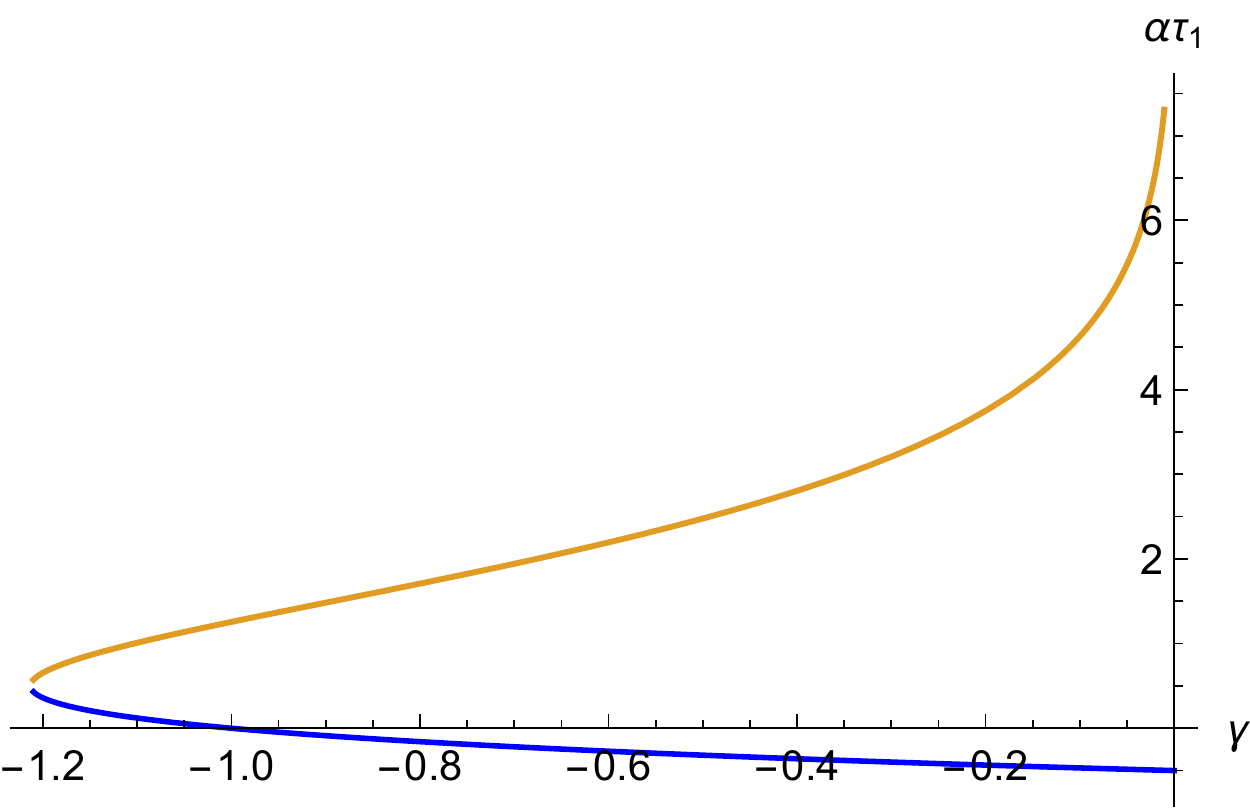}
	\caption{Plot of solution (\ref{tau1fix}) for $\alpha \tau_1$ 
	as a function of the ratio $\gamma=\frac{{\cal W}_0}{A}$.  The
	 orange (upper)  and blue (lower) curves represent the $W_{-1}$  and $W_0$  branches, respectively. Acceptable values ($a\tau_1>0$) for the blue curve are compatible only with its section satisfying $\gamma <-1$.} 
	\label{fL1}
\end{figure}

Taking into account the above corrections,  the  K\"ahler potential takes the form
\begin{equation}
{ \cal  K}=
-2\ln\left(\sqrt{{\tau}_1{\tau}_2{\tau}_3}+{\xi}+{ \eta} \ln{ ({\tau}_1{\tau}_2{\tau}_3)}\right)\equiv 
-2\ln\left({\cal V}+{\xi}+{\eta} \ln {\cal V}\right)~\cdot \label{KalCor}
\end{equation}
The covariant derivative of the superpotential 
w.r.t. the K\"ahler modulus $\rho_1$ is defined in the usual manner, i.e., $D_{\rho_1}W= \partial_{\rho_1} {\cal W}+{\cal W}\partial_{\rho_1} {\cal K}$. 
 Working in the large volume limit,  terms  proportional to $\xi$ and $\eta$ coefficients compared to the volume ${\cal V}$  are ignored. 
Writing the K\"ahler potential as
 \begin{align}
\mathcal{K}=-2\log(\sqrt{(\rho_1-\bar{\rho}_1)(\rho_2-\bar{\rho}_2)(\rho_3-\bar{\rho}_3)}+\mathcal{O}(\eta,\xi))
 \end{align}
and taking the derivatives~\footnote{We denote with calligraphic letters ${\cal W}_0, {\cal W}$ 
	the tree-level and corrected  superpotential and reserve the
	symbols ${W}, {W_0}, {W_{-1}}$ for the Lambert W-function.}  
 \begin{align}
 \partial_{\rho_1}\mathcal{W}= i\alpha A e^{i \alpha \rho_1},\;\; \partial_{\rho_1}\mathcal{K}=-\dfrac{1}{\rho_1-\bar{\rho}_1}
 \end{align}
it is readily found that
\be 
\left. D_{\rho_1}{\cal W}\right|_{\rho_1=i\tau_1}= i e^{-\alpha\tau_1}\left(\alpha A+\frac{A+{\cal  W}_0e^ {\alpha\tau_1}}{2\tau_1}\right)~\cdot  \label{CovDer0}
\ee 
The corresponding supersymmetric condition, 
$D_{\rho_1}{\cal  W}=0$, fixes the value of the modulus $\tau_1={\rm Im}{\rho_1}$ in terms 
of the tree-level superpotential ${\cal W}_0$ (determined by the choice of the fluxes) and  the coefficients $\alpha, A$ - related to non-perturbative contributions.  Thus, the vanishing of the derivative (\ref{CovDer0})  yields
\ba
\tau_1 =-\frac{1+2 w}{2\alpha }~, \label{tau1fix} 
\ea
where $w$ represents either of the two branches $W_{0}, W_{-1}$, of the Lambert W-function 
\ba 
w&=&W_{0/-1}({\tiny \frac{\gamma}{2\sqrt{e}}})~.\label{lambert} 
\ea 
In (\ref{lambert}), the convenient definition has been introduced 
\ba 
\gamma&=&\frac{{\cal W}_0}{A}~\cdot \label{gamma}
\ea
 Real values of the solution are compatible  with the bound $\gamma \ge -2e^{-1/2}\approx -1.213$ for both branches.
  For the ``lower'' branch $W_0$, equation (\ref{tau1fix}) implies the constraint 
$\alpha \tau_1 \le 1/2$. Requiring also $\alpha \tau_1 >0$
it is found that the ratio  $\gamma ={\cal W}_0/A$ is confined in the region:
\be    
-1.213\lsim \gamma\le -1~\cdot \label{LowRange}
\ee
This solution is depicted with the
blue  curve in  figure~\ref{fL1}.
The corresponding regions for  the ``higher'' branch $W_{-1}$,
depicted with the orange curve in figure~\ref{fL1}, are
\be 
-1.213\lsim \gamma\le 0~,\label{HighRange}
\ee 
and $\alpha \tau_1 \in [\frac 12,\infty]$ .
\subsection{F-term potential with quantum corrections}

 The F-term scalar potential is computed by inserting (\ref{KahlerP}) into (\ref{KahlerV}). This yields a
rather complicated formula  which is not very illuminating, however,  in the large volume 
limit it suffices to expand it w.r.t. the
small parameters  $\eta$ and $\xi/{\cal V}$ and obtain a simplified  form. Thus,
 without loosing 
its essential features, in this approximation  the  potential is written as a sum of three parts, as follows:
\be 
V_F\approx V_{F_1}+ V_{F_2}+ V_{F_3}~\cdot 
\label{VFappr} 
\ee 
The various parts of the RHS in~(\ref{VFappr}) are given by
\be
\begin{split}
	\label{3partVeff} 
	V_{F_1}&= \frac 32 {\cal W}_0^2\; \frac{\xi -2\eta ( 4- \ln{\cal V})}{{\cal V}^3}-9{\cal W}_0^2\frac{ \xi \eta  \log ({\cal V})}{{\cal V}^4}~,\\
		V_{F_2}&=4 \frac{\alpha\tau_1 }{{\cal V}^2}\tilde A(\tilde A+a\tau_1 \tilde A+{\cal W}_0)~,\\
	V_{F_3}&=\tilde A( \tilde A\,f+{\cal W}_0\,g)
~\cdot 
\end{split}
\ee 
where  $\tilde A=   e^{- \alpha \tau_1} A$ and  ${\cal O}(\frac{1}{{\cal V}^5})$  or higher terms in the expansion are ignored. Also
\begin{align*}
f=&\dfrac{3 \xi-8 \eta  (2 \alpha  \tau_1  (2 \alpha  \tau_1 +3)+3) -4 \xi\alpha    \tau_1 (\alpha  \tau_1 +1)-2 \eta  (2 \alpha  \tau_1 -1) (2 \alpha  \tau_1 +3) \log \mathcal{V}}{2\mathcal{V}^3}\\
&+\dfrac{ \eta  \xi  (2 \alpha  \tau_1+3) ((6 \alpha  \tau_1 -3) \log\mathcal{V}-4 \alpha  \tau_1 )}{\mathcal{V}^4},\\
g=&\dfrac{(3 -2 \alpha\tau_1)\left(\xi+2\eta\log(\mathcal{V})\right)-24 \eta(1+\alpha\tau_1)}{\mathcal{V}^3}-6\eta \xi\dfrac{\left(3-2\alpha \tau_1\right)\log\mathcal{V}+2\alpha\tau_1}{\mathcal{V}^4}~\cdot
\end{align*}

In the above  all three K\"ahler moduli  $\tau_i$ are  expressed in terms of the volume ${\cal V}$ with $\tau_1$ being considered at its critical value $\tau_1^{cr}$ given in (\ref{tau1fix}),  fixed from the supersymmetric conditions imposed on  the superpotential. Therefore,  only the two of them, namely  $\tau_2$ and $\tau_3$ are left undetermined which 
appear only in the combination
 $\tau_2\tau_3={\cal V}/{\tau_1^{cr}}$~\footnote{From now on, we drop ``${cr}$'' from $\tau_1^{cr}$ and write just $\tau_1$ for simplicity.}.
 It is to be noted that, since there are regions of solutions $D_{\rho_1}{\cal  W}=0$,
 	where $\tau_1$ is hierarchically smaller than the rest 
 	of the moduli and $\alpha \tau_1$ receives relatively moderate values, (see figure~\ref{fL1}),  terms involving  $\tilde A^2$ have also been retained. 
 	It should be further pointed out that, in principle, there are regions of the parameter space (in particular those with large values of $\tau_1^{cr}$) where such 	terms are comparable to ${\cal V}^{-5}$, the latter being omitted  in the large volume expansion. Then, 
 	$\tilde A^2$ terms  could be safely neglected too. In this case  non-perturbative corrections are 
 	suppressed and  the perturbative logarithmic corrections prevail. One of the objectives of  this work, however, is to also probe regions where all terms of
 	(\ref{3partVeff}) have comparable contributions to the total potential
 in~(\ref{VFappr}).

At this point, it is worth clarifying the origin of the components~(\ref{3partVeff}). The term $V_{F_1}$ is derived  from the $\alpha'$ and perturbative
string loop corrections  due to the localised EH terms, both entering in the K\"ahler potential (\ref{KalCor}).
 Indeed, switching off the non-perturbative corrections, i.e. setting $A=0$, the only term
remaining in  (\ref{3partVeff}) is the $V_{F_1}$ component which is identified with the one given in~\cite{Antoniadis:2018hqy} where only perturbative corrections are studied.  Setting $\eta$ and $\xi$ equal to zero, 
the only term that remains is the second component, $V_{F_2}$. This  
contribution comes exclusively from the non-perturbative corrections which were included in the  superpotential. Finally, the third component $V_{F_3}$ 
is a mixing term and it is non-vanishing  only when both perturbative and non-perturbative  corrections 
are present.\\
\noindent
    As an additional check with regard to the non-perturbative part,  the appropriate  limit of (\ref{3partVeff}) is taken to  reproduce the already  known results in the literature~\cite{Kachru:2003aw}. Indeed, for   $\eta=\xi= 0$  the  scalar potential   becomes
\begin{equation}\label{eq:KKLT}
V_{F_2}=\dfrac{4e^{-2\alpha \tau_1}\alpha A}{\tau_2\tau_3}(e^{\alpha\tau_1}{\cal  W}_0+A+\alpha\tau_1 A)~\cdot 
\end{equation}
Solving \eqref{CovDer0} w.r.t. the ${\cal  W}_0$, it is found that:
\begin{equation}
{\cal  W}_0=-Ae^{-\alpha \tau_1}(1+2\alpha \tau_1)~\cdot 
\end{equation}
Substituting in \eqref{eq:KKLT}  while putting $\tau_3\rightarrow \tau,\tau_2\rightarrow \tau,\tau_1\rightarrow \tau$  the  result is
\begin{equation}
V_{min}=-\dfrac{4e^{-2\alpha\tau}\alpha^2 A^2}{\tau}~,
\end{equation}
which (up to numerical factor related to the multiplicity of the K\"aher moduli) coincides with the solution of ~\cite{Kachru:2003aw}.

To proceed with  the minimisation of the scalar potential  (\ref{3partVeff}), a more convenient form will be worked out. To
this end,  the following parameter is  introduced
\ba 
\epsilon &=& \frac{2 w+1}{w}~\cdot \label{Epsilon} 
\ea 
Furthermore, for later convenience, 
the range of the various parameters defined up to this point for the two branches of the solution are shown in Table~\ref{ranges}. 
As already noted, in the LVS regime  it would be more suitable to have large directions given by the lower branch $W_{-1}$. 
However, these solutions represent instanton corrections, and as it is obvious, the $W_0$ branch is a strongly coupled region, where higher order corrections should be taken into account. For the reasons discussed above and for the correctness of the effective approach, from now on only the $W_{-1}$ branch 
will be  considered as the solution for the $\tau_1$ modulus. \footnote{The current understanding of the non-perturbative physics prevent a complete study of the other branch. A way of treating instanton corrections from $D3$-branes is presented in~\cite{Baumann:2010sx}.}

\begin{table}
	\begin{center}  \small%
		\begin{tabular}{|p{2.5cm}|p{1.0cm}|p{1.2cm}|p{1cm}|p{1.0cm}| p{1.1cm}|}
			\hline
			Branch & $a\tau$ & $\gamma$ & $w$&$\epsilon$ \\
			\hline
			$w=W_0(\frac{\gamma}{2\sqrt e})$ & $0$ & $-1$ &$-\frac 12$ &0\\
			& $\frac 12$ & $-\frac{2}{\sqrt e}$ &$-1$ &1\\
			\hline
			\hline
			$w=W_{-1}(\frac{\gamma}{2\sqrt e})$ & $\infty$ & $0$ &$-\infty$ &2\\
			& $\frac 12$ & $-\frac{2}{\sqrt e}$ &$-1$ &1\\
			\hline
		\end{tabular}
	\end{center}
	\caption{The range of the various parameters used in the analysis. }
	\label{ranges}
\end{table}

\noindent
Using the above definitions, and the identities $2 w=\gamma e^{\alpha\tau_1}=-(2\alpha\tau_1 +1)$ resulting from (\ref{tau1fix}-\ref{gamma})
 the F-term potential~(\ref{VFappr}) can be cast in a convenient compact form.
Considering  the $V_{F_2}$ piece in particular, under successive substitutions of  $\gamma=\frac{{\cal W}_0}{A}$, 
$2\alpha \tau_1=-(1+2w)$ and  $\gamma e^{\alpha \tau_1}=2w$ its 
third term gives 
 \begin{align}
 	\begin{split}
 		&\dfrac{4\alpha \tau_1 A \mathcal{W}_0e^{-\alpha \tau_1}}{\mathcal{V}^2}=-2\dfrac{\mathcal{W}_0^2(1+2w)}{\gamma e^{\alpha \tau_1}\mathcal{V}^2}=-\dfrac{\mathcal{W}_0^2(1+2w)}{w\mathcal{V}^2}
 	\end{split}
\end{align}
Continuing as above, it is found that all three terms of $V_{F_2}$ add up to:
$$V_{F_2}=-\dfrac{\mathcal{W}_0^2}{\mathcal{V}^2} \dfrac{(1+2w)^2}{4w^2}=-\frac{(\eps \mathcal{W}_0)^2}{4 \mathcal{V}^2}~.$$
Finally, 
 the following compact form  of the whole $V_F$ potential is obtained
\be 
V_{F}\approx \left( {\epsilon \cal W}_0\right)^2\left(
\frac{ 2 \xi-{\cal V} +4 \eta  (\log
	({\cal V})-1)}{4 {\cal V}^3}-\eta\xi \frac{3\log({\cal V})-1 }{{\cal V}^4}
\right)+{\cal O}\left(\frac{1}{{\cal V}^5}\right)~\cdot 
\label{VFc}
\ee
In the present approximation, valid in the large volume limit, 
it is observed that the parameters associated with the
 non-perturbative effects appear in the F-term
potential as an overall positive-definite factor 
$\epsilon^2$ where $\epsilon$ is defined in~(\ref{Epsilon}). 
Thus, the shape of $V_F$ is  controlled by the second factor 
which exhibits  the volume dependence and involves 
 the parameters
$\xi$ and $\eta$ coming from the perturbative corrections 
in the K\"ahler potential.  Indisputably, the properties of 
the potential depend decisively on the signs of $\xi, \eta$ given in~(\ref{etaxi})
which convey topological and geometric information of the compactification manifold.
For closed orientable smooth manifolds 
and the particular  D7-branes  set up~\cite{Antoniadis:2019rkh} in the present study 
the choice $\chi<0, \xi>0$ will be adopted. Then, 
 dropping the subleading terms of order $\propto \frac{1}{{\cal V}^4}$  and higher in the large volume regime,  and  requiring 
the vanishing of the first derivative,  it is 
found that the volume at the minimum of the
potential is given by 
\be 
{\cal V}_{\rm min}=-6{\eta}\,W_{0}\,\left(-\frac{1}{6\eta}
e^{\frac{4}{3}-\frac{\xi}{2\eta}}\right)~, \label{Volmin}
\ee 
where $W_{0}$ is the Lambert W-function. 
  Substituting  ${\cal V}_{\rm min}$ into the second derivative yields:
\begin{equation}
\dfrac{d^2V_F}{d\mathcal{V}^2}=(\epsilon { \cal W}_0)^2\dfrac{\mathcal{V}-6\eta}{2\mathcal{V}^5}~\cdot
\end{equation}
Hence, a minimum exists as long as $\mathcal{V}\ge 6\eta$ which is 
obviously true in the large volume regime, although this corresponds to an AdS vacuum.   
Nonetheless,  this  can be  naturally uplifted to a dS minimum, when 
D-term contributions  are taken into account. It should be pointed out too, that
minimisation of $V_F$ w.r.t. ${\cal V}$  stabilises only the combination $\tau_2\tau_3={\cal V}/\tau_1$ leaving 
another independent combination of $\tau_2, \tau_3$  moduli  undetermined. This will also be rendered 
with the inclusion of the  D-terms  in the next section.

\section{De Sitter spacetime from D-terms}

Various mechanisms have been proposed to uplift a supersymmetric AdS vacuum and obtain a metastable dS minimum and thusly, a  positive 
cosmological constant. Amongst others, these include $\overline{D3}$ branes, and nilpotent superfields~\cite{Ferrara:2014kva}. 
In~\cite{Antoniadis:2018hqy} it was proven that  it is sufficient to consider  the D-term contributions associated with $U(1)$ factors
 which arise in the presence of  the intersecting $D7$  branes already  included in the geometric configuration. 
Flux generated  D-terms   have the general form~\cite{Burgess:2003ic,Haack:2006cy}
 \be 
 V_{D}= \frac{g_{D7_i}^{2}}{2}\left(Q_i\partial_{\rho_i} K+\sum_j q_j |\Phi_j|^2\right)^2,\; \frac{1}{g_{D7_i}^{2}}={\rm Im}\rho_i+\cdots 
 \ee 
 where $Q_i,q_j$ are ``charges'' and  $\{\cdots\}$ stand for flux and dilaton dependent corrections while the $\Phi_j$  fields  depend on the 
 specific field theory model.
 For zero $\Phi_j$-vevs  the model dependent term vanishes 
  (see discussion in~\cite{Haack:2006cy}) and it turns out that $	V_{D}\approx  Q_i^2/\tau_i^3$. 
 Then, the generic form of the corresponding D-term potential can be approximated by~\cite{Antoniadis:2018hqy}
 \be 
{V}_{ \cal D}\; = \;\sum_{i=1}^3\frac{d_i}{\tau_i} \left(\frac{\partial { {\cal K}}}{\partial {\tau_i}}\right)^2
\;\approx \;
\sum_{i=1}^3 \frac{d_i}{\tau_i^3}\;\equiv\; \frac{d_1}{\tau_1^3}+\frac{d_3}{\tau_3^3}+\frac{d_2\tau_1^3\tau_3^3}{{\cal V}^6}~,\label{VDappr}
 \ee 
 where,  $d_i\approx Q_i^2>0$ and in the last expression  the modulus $\tau_2$  has been traded
 with the internal volume modulus ${\cal V}$, i.e.,  $\tau_2={\cal V}^2/(\tau_1 \tau_3)$. 
Thus, the effective potential being the sum of the (\ref{VFc}) and (\ref{VDappr}),  $V_{\rm eff}= V_F+V_D$, is given as a function 
 of  $\tau_1,\, \tau_3$ and ${\cal V}$:
\begin{equation}
V_{\rm eff}\approx -\left( {\epsilon \cal W}_0\right)^2\,
\frac{{\cal V}- 2 \xi +4 \eta  (1-\log
	({\cal V}))}{4 {\cal V}^3}+\frac{d_1}{\tau_1^3}+\frac{d_3}{\tau_3^3}+\frac{d_2\tau_1^3\tau_3^3}{{\cal V}^6}~\cdot 
\label{VFD}
\end{equation}

 Two different ways shall be considered  in incorporating 
 the D-terms in the minimisation procedure.  In the first case, 
 the modulus  $\tau_1$ will be treated as constant with its value fixed
 by the condition (\ref{tau1fix}). Then, the potential $V_{\rm eff}$ will be considered
 as a function  only of the variables  ${\tau_2}$ and $\tau_3$, or equivalently ${\cal V}$ and $\tau_3$.
 Hence, since  $\tau_1$ is treated as constant, the minimisation of  $V_{\rm eff}$ will be 
 worked out only w.r.t.  ${\cal V}$ and $\tau_3$. This treatment might not be completely justified  and
 probably one has to rely on physics beyond the  present context~\footnote{ For example, $\tau_1$  could be perceived as  
 	the inter-brane separation   where its value is	already fixed in connection to the modified Randall-Sundrum model proposed 
 	in the literature  (e.g.~\cite{Brummer:2005sh,Randall:2019ent}).}, 
notwithstanding, it is envisaged that this case is worth   exploring since 
the resulting potential exhibits interesting features for cosmological applications.  In the second approach, a fully fledged minimisation
procedure is performed w.r.t. all three moduli $\tau_{1,2,3}$ (or 
 $\tau_{2,3}$ and ${\cal V}$).

 ${\cal A})\,$
 Starting with the first approach, it is assumed  that the $\tau_1$ modulus acquires 
 its critical value already determined by the condition (\ref{tau1fix}). Since $d_1>0$, the corresponding D-term 
  automatically  operates as a natural uplift mechanism of the potential, $V_{up}\equiv \frac{d_1}{\tau_1^3}$.
Therefore,  in addition to the volume there is only one more variable left,  namely the modulus $\tau_3$. Then, the   
minimisation condition yields
\be 
\tau_3=\left(\frac{d_3}{d_2}\right)^{1/6}\frac{\cal V}{\tau_1^{1/2}} ~\cdot 
\ee 
Substituting this  into (\ref{VFD}) and 
 minimising the resulting effective potential  w.r.t.  the volume, it is found that  \\
\begin{equation}
V_{\rm eff}|_{\tau^{min}_3}=-(\epsilon {\cal W}_0)^2\dfrac{\mathcal{V}-2\xi+4\eta(1-\log\mathcal{V})}{4\mathcal{V}^3}
+\dfrac{2d\tau_1^{3/2}}{\mathcal{V}^3}
+\dfrac{d_1}{\tau_1^3}~\cdot 
\label{Veffin}
\end{equation}
where $d=\sqrt{d_2d_3}$. As can be seen, the term proportional to $d$ can be absorbed in a redefinition of $\xi$. 
As a result, the form of  $V_{\rm eff}$  is just that of the F-term potential supplemented by the uplift term $d_1/{\tau_1^3}$.
Defining the new parameters 
$$\zeta=\xi+4\delta,\; \delta=\dfrac{d\tau_1^{3/2}}{(\epsilon \mathcal{W}_0)^2} ~, $$
the volume at the minimum is given now by
\be 
{\cal V}_{V'=0}=-6{\eta}\,W_{0}\,\left(-\frac{1}{6\eta}
e^{\frac{4}{3}-\frac{\zeta}{2\eta}}\right)~.
\label{Volextrem}
\ee 

\par  The value of the potential at the minimum is obtained by substituting  (\ref{Volextrem})  into (\ref{Veffin}).
Requiring a dS solution, the following  bound for the uplift parameter $d_1$
is derived 
\begin{equation}\label{uplift}
d_1\,>\,(\epsilon \mathcal{W}_0)^2\tau_1^3\dfrac{{\cal V}_{min}-4\eta}{12{\cal V}_{min}^3}\approx 
(\epsilon \mathcal{W}_0)^2\dfrac{\tau_1^3}{12{\cal V}_{min}^2}~\cdot
\end{equation}
Clearly, in the large volume regime, this is easily satisfied 
even for  small values of the  coefficient $d_1$ as long as $\tau_1$ 
remains finite.

\begin{figure}[H]
	\centering
    \includegraphics[scale=0.88]{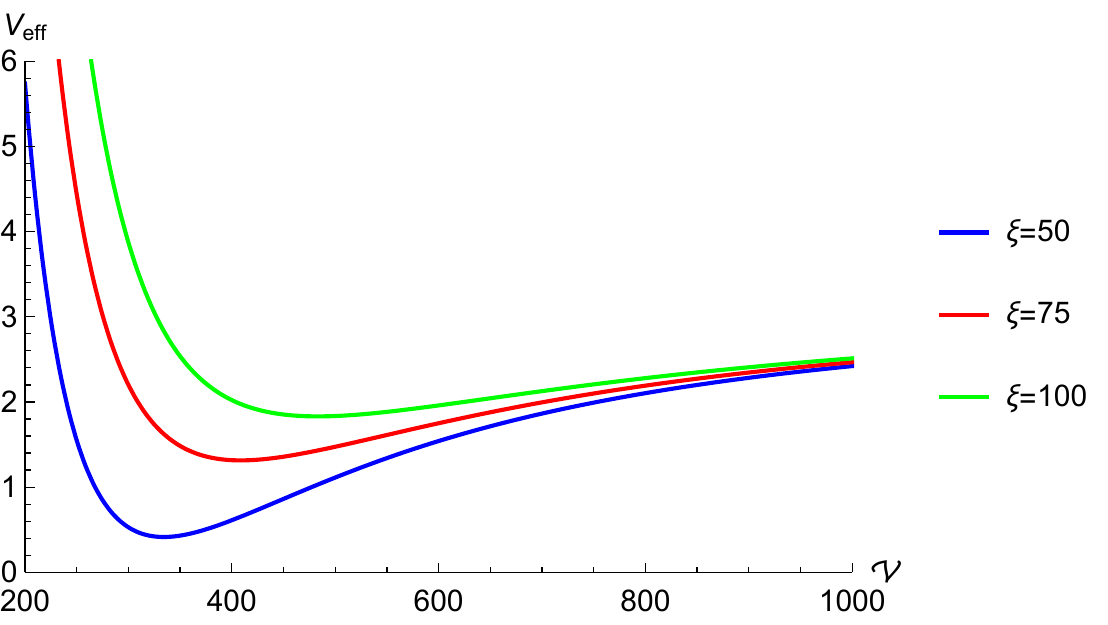}
	\caption{Plot of $V_{\rm eff}\times 10^6$ potential vs the volume modulus  for  $\epsilon \mathcal{W}_0=1.9,\tau_1=40,\eta=-1/2$, $d=2.35 \times 10^{-1}, d_1=0.2$ and	three values of the topological parameter $\xi$. } 
	\label{figVFD1}
\end{figure}

\begin{figure}[H]
	\centering
	\includegraphics[scale=0.88]{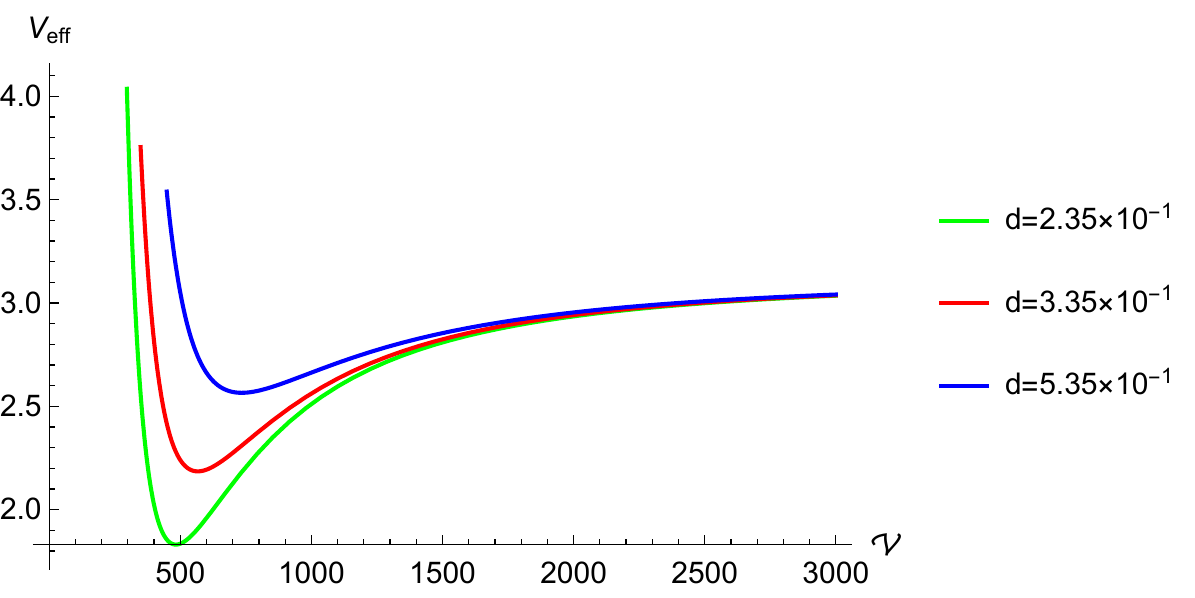}
	\caption{Plot of $V_{\rm eff}\times 10^6$ potential vs the volume modulus for  $\epsilon \mathcal{W}_0=1.9,\tau_1=40,\eta=-1/2$, $\xi=100, d_1=0.2$ for	three values of the D-term coefficient $d$. } 
	\label{figVFD2}
\end{figure}

\begin{figure}[H]
	\centering
	\includegraphics[scale=0.5]{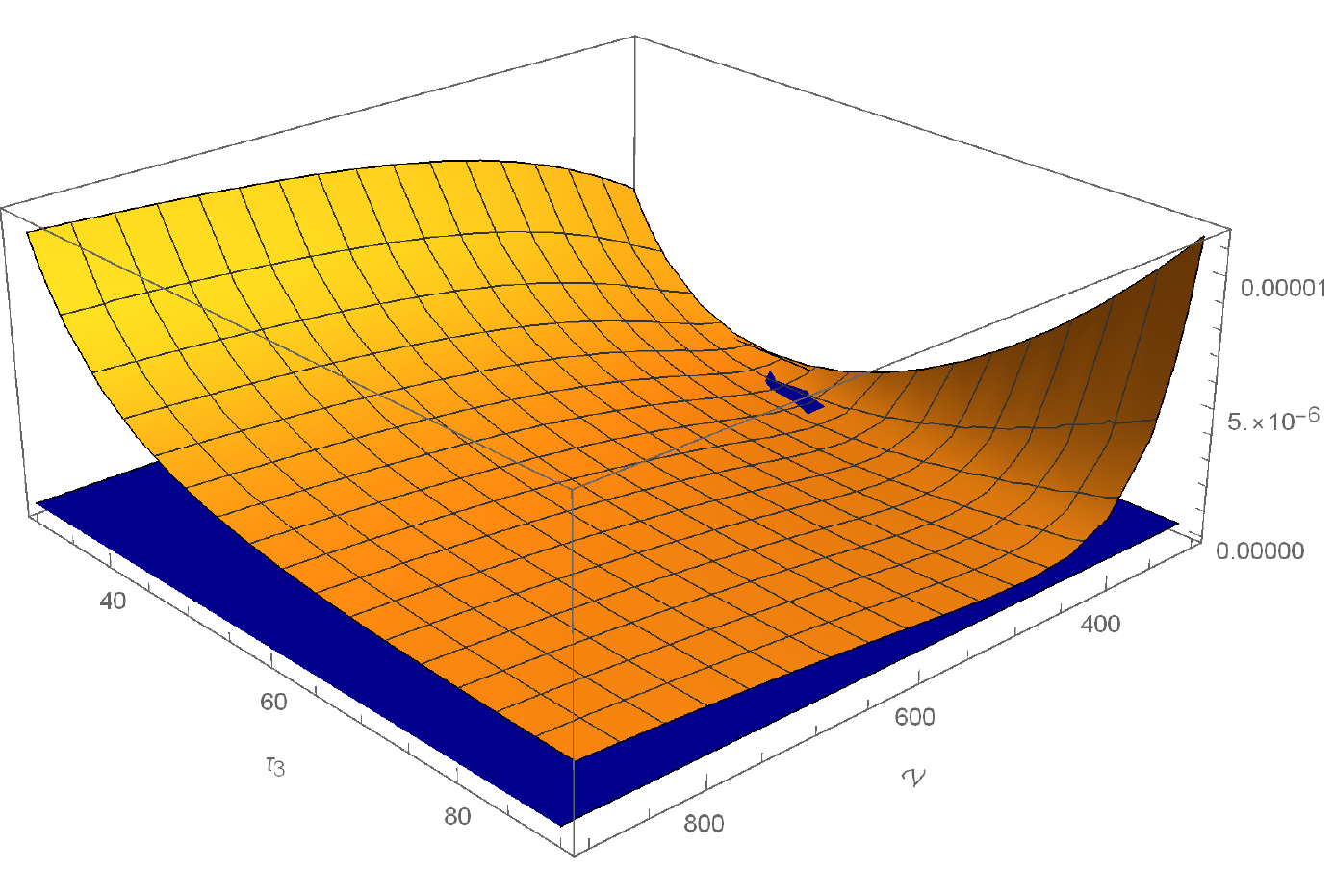}\; \includegraphics[scale=0.34]{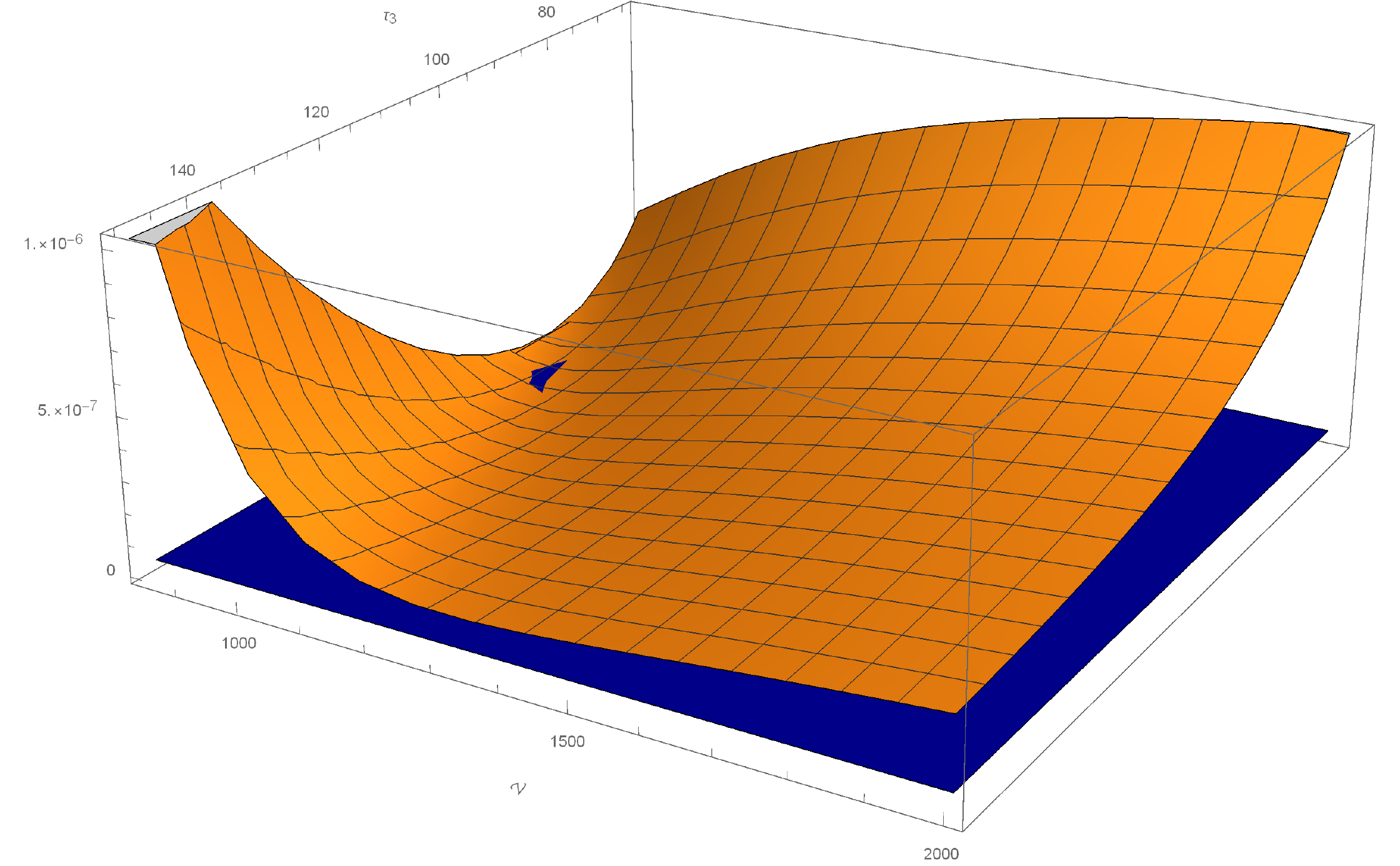}
	\caption{Left panel: A 3D Plot of $V_{\rm eff}$  vs the moduli ${\cal V}, \tau_3$ where its dS minimum is displayed for the parameter values $\epsilon\mathcal{W}_0=1.9,\xi=50,\tau_1=40,d_2=d_3=0.235,\eta=-0.5,d_1=0.2$. A horizontal blue plane has been  drawn at a tiny height  above the $V_{eff}^{min}$ value, intersecting $V_{\rm eff}$ at the vicinity of its  minimum (blue dot). Right panel: A similar plot of $V_{\rm eff}$ is shown for $d_1=0.4$ and $\tau_1=100$.  } 
	\label{V3D}
\end{figure}

In figures~\ref{figVFD1} and \ref{figVFD2}  
plots of ${ V_{\rm eff}}$  vs the volume modulus ${\cal V}$ are shown for various values  of the
parameters obeying the appropriate constraints discussed in 
the analysis.  The two plots of figure~\ref{V3D} show the dS minimum of the potential 
vs $\tau_3$ and the volume modulus.  

\par At the level of the effective model,  the properties  of the scalar potential  have significant  phenomenological and  cosmological implications.
In this respect, the shape of $V_{\rm eff}$ as shown in figure~\ref{figVFD2}  looks promising in accommodating slow roll  cosmological inflation
and it is a theme worth contemplating in a future publication.

${\cal B})\,$ Next,  an alternative approach is considered, where the 
potential is minimised w.r.t. both  moduli $\tau_{1,3}$ as well as
the volume ${\cal V}$. Proceeding as in~\cite{Antoniadis:2018hqy},
it is found that the $\tau_{1,3}$ moduli are stabilised at 
\be 
\tau_k =\,\left(\frac{d_k}{d}{\cal V}^{2}\right)^{1/3},\;k=1,3,\; {\rm where}\; 
d=(d_1d_2d_3)^{\frac 13}\label{tau13}
\ee



\noindent 
The potential takes the form

\begin{equation}
V_{\rm eff}|_{\tau^{min}_{1,3}}=-(\epsilon {\cal W}_0)^2\dfrac{\mathcal{V}-2\xi+4\eta(1-\log\mathcal{V})}{4\mathcal{V}^3}
+\dfrac{3d}{\mathcal{V}^2}
\label{Veffin2}
\end{equation}

\begin{figure}[H]
	\centering
	\includegraphics[scale=0.9]{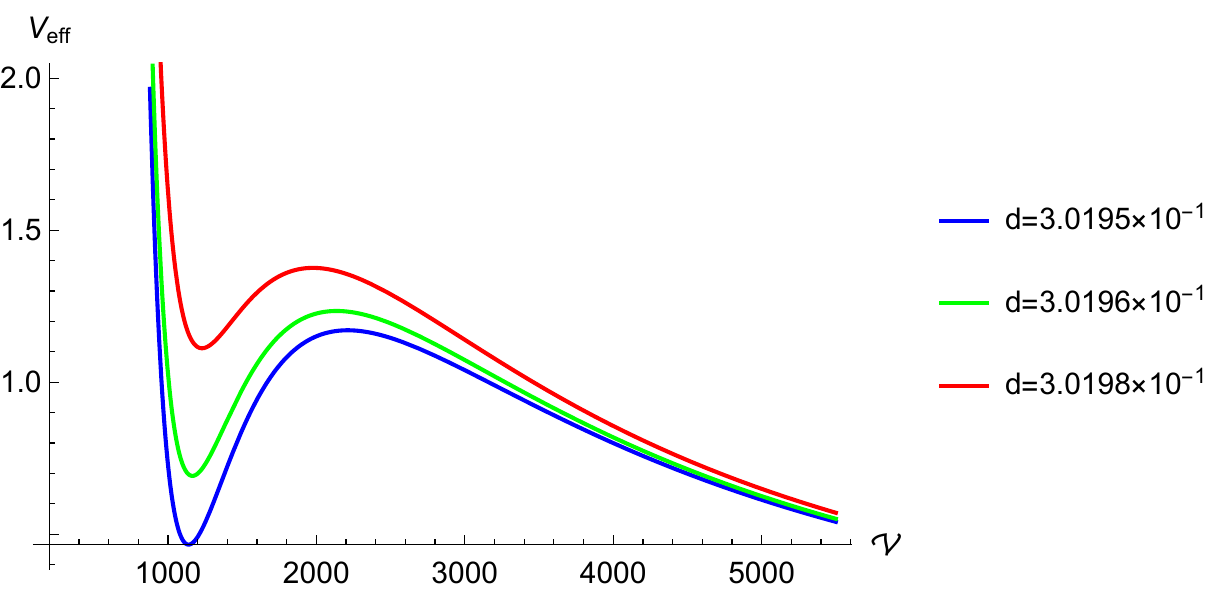}
	\caption{Plot of $V_{eff}\times 10^{10}$ potential vs the volume modulus for  $\epsilon \mathcal{W}_0=1.9,\eta=-1$, $\xi=10$ for three values of the D-term coefficient $d$. } 
	\label{figure5}
	\end{figure}

\noindent 
At the minimum of the potential the volume modulus takes the value
\begin{align}
\mathcal{V}_{V'=0}=\dfrac{6|\eta|}{1-12 \mathtt{r}}W_0\left(\dfrac{1-12 \mathtt{r}}{6|\eta|}e^{\frac{4}{3}-\frac{\xi}{2\eta}}\right)\label{Veffmin2}
\end{align}
As in the previous case, the following two constraints are imposed: i) the 
argument of the $W_0$ function must be larger than $-1/e$ and ii) the
potential at the minimum must be positive. Once these restrictions
are implemented, 
the ratio   $\mathtt{r}=\frac{d}{(\epsilon{\cal W}_0)^2}$ of the $F$- and $D$-term  coefficients 
is found to be bounded in the region 
\be 
\frac{1}{12}-\frac{\eta}{3{\cal V}_{min}}\le \mathtt{r} \le 
\frac{1}{12}-\frac{\eta}{2}
e^{\frac{\xi}{2\eta}-\frac{7}{3}},\;\;
\label{bounds}
\ee 
For large volumes, the above bounds allow only a tiny  region  in the vicinity of  $1/12~$.
Given the ratio $\mathtt{r}$, the inequalities~(\ref{bounds})  imply also an upper bound on $\xi$:
\begin{align}
-\dfrac{\eta}{3\mathcal{V}_{min}}<-\dfrac{\eta}{2}e^{\frac{\xi}{2\eta}-\frac{7}{3}}&\Rightarrow 
\xi < 2|\eta|\left( \ln\frac{6|\eta|}{12 \mathtt{r}-1}-\frac 73\right) 
\end{align}
In figure~\ref{figure5} the potential is plotted vs the volume for
the set of parameters $\epsilon {\cal W}_0=1.9, \xi=10, \eta=-1$ and three 
values of the D-term coefficient $d$. A dS minimum is obtained for a very short range of $d$. In contrast to the first case, here there is no constant uplift term, all $V_{\rm eff}$ terms are suppressed by powers of
${\cal V}$ and the potential asymptotically approaches  zero as ${\cal V}\to \infty$. 

\section{Conclusions }

In this letter  the issues of K\"ahler moduli stabilisation and   the existence  of de Sitter spacetime vacua  in 
the string landscape have been revisited within the framework of type IIB string  theory. 
The analysis has been conducted in the context of a geometric background of three intersecting seven-branes 
and three K\"ahler moduli  $\tau_{1,2,3}$  associated with the large codimension-two volumes transverse to each one of them. More precisely the
implications of the combined effects of  the well-known non-perturbative contributions 
in the fluxed superpotential and the recently studied perturbative 
logarithmic corrections~\cite{Antoniadis:2019rkh} in  the K\"ahler potential have been considered. 
Regarding the former corrections, only  one modulus  $\tau_1$ is taken into account in the present study. 
The origin of the latter come from string loop corrections 
when closed strings emitted from localised Einstein-Hilbert terms, $R_{(4)}$,
traverse the codimension-two volume towards the seven-brane probes.
 Such $R_{(4)}$ terms come from the $R^4$ corrections of the 
 effective ten-dimensional string  action and, remarkably, appear only in four spacetime dimensions. 

As in the previous studies with perturbative corrections~\cite{Antoniadis:2018hqy}, de Sitter vacua are possible  only due to  D-terms  
originating from intersecting $D7$-branes and their associated $U(1)$ factors.
For finite values of the volume modulus the general shape of the effective potential $V_{\rm eff}$  near the minimum  is similar  with those found in 
the study including only perturbative corrections.  A substantial difference, however,  arises in their asymptotic behavior. 
In the  perturbative case the dS potential possesses a minimum  at some finite $ {\cal V}_{min}$  as well 
as a (local) maximum at a finite value $ {\cal V}_{max}>{\cal V}_{min} $ before it approaches zero at ${\cal V}\to \infty$. On the
contrary,  for the case studied in this work, for values ${\cal V}> {\cal V}_{min}$  the potential approaches asymptoticly this 
maximum at ${\cal V}\to \infty$.  This  is due to the fact that the value of the K\"ahler  modulus $\tau_1$ appearing   in the 
non-perturbative superpotential  is  fixed by the supersymmetric conditions. Consequently, the corresponding  D-term  enacts as a 
constant uplift $V_{up}$ turning an AdS vacuum to a dS one, similar to the Fayet-Iliopoulos  term considered in a supergravity  context~\cite{Cribiori:2017laj,Antoniadis:2019hbu}. 
Furthermore, a separate study is conducted where the 
modulus $\tau_1$ is also treated dynamically in the
minimisation of   the scalar potential. In this case  
both  $ {\cal V}_{max}$ and ${\cal V}_{min} $ are finite 
and $V_{\rm eff}$ exhibits a behaviour similar to that
studied in~\cite{Antoniadis:2018hqy}.

As has been already emphasised, the above interesting
features  have been  demonstrated for a small number of K\"ahler moduli
while many CY manifolds predict large numbers of them. Some recent
attempts of the authors of~\cite{Burgess:2020qsc}  have extended the analysis on arbitrary numbers, however,
effects of logarithmic  corrections considered here are not captured
by their approach. 
It is envisaged that the present approach is a starting point in aspiring to 
more complicated  cases where  large numbers of moduli and    various types of quantum corrections are involved in these
computations. Furthermore, due to the variant  features of the effective
potential as compared to previous studies, novel
cosmological implications are  anticipated  worth exploring in future work.

\vspace{2cm}

{\bf Acknowledgement}
{\it ``This research work was supported by the Hellenic Foundation for
	Research and Innovation (H.F.R.I.) under the ``First Call for
	H.F.R.I. Research Projects to support Faculty members and
	Researchers and the procurement of high-cost research equipment
	grant'' (Project Number: 2251)''}.  {\it The authors would like to thank
A. Karozas for comments on an earlier version of the manuscript.}

\end{document}